\documentclass[12pt,a4paper,twoside]{article}

\usepackage[cp1251]{inputenc}
\usepackage[T2A]{fontenc}
\usepackage[russian]{babel}

\usepackage[dvips]{graphicx}
\usepackage{amsmath}
\usepackage{amssymb}
\usepackage{amsxtra}
\usepackage{amsthm}
\usepackage{latexsym}
\usepackage{array}
\usepackage{multirow}
\usepackage{hhline}

\newcommand{\bs}{\boldsymbol}
\newcommand{\la}{{\lambda}}
\newcommand{\ri}{{\rm i}\,}
\newcommand {\mP}{\mathcal{P}}
\newcommand {\bbR}{\mathbb{R}}
\newcommand {\bbI}{\bbR^3(h,g,k)}

\newcommand {\sil}{\mathcal{S}_\ell}
\newcommand {\sila}{\mathcal{S}_\ell(\lambda)}
\newcommand {\wsi}{\Pi}

\newcommand {\wsa}{\Pi_2}
\newcommand {\mm}{\mathcal{M}_2}

\newcommand {\wsb}{\Pi_3}
\newcommand {\mn}{\mathcal{M}_3}

\newcommand {\wsc}{\Pi_1}
\newcommand {\mo}{\mathcal{M}_1}
\newcommand {\gaa} {\gamma_1}
\newcommand {\gab} {\gamma_2}
\newcommand {\gac} {\gamma_3}
\newcommand {\gan} {\delta_0}
\newcommand {\htan} {h_{\rm tan}}
\newcommand {\Fun} {Z}

\newcommand{\cons}{\mathop{\rm const}\nolimits}
\renewcommand{\cosh}{\mathop{\rm ch}\nolimits}
\renewcommand{\sinh}{\mathop{\rm sh}\nolimits}
\newcommand {\vk}{\varkappa}
\newcommand {\ro}{\rho}
\newcommand {\ds}{\displaystyle}
\newcommand{\gs}{\geqslant}
\newcommand{\ls}{\leqslant}
\newcommand{\sgn}{\mathop{\rm sgn}\nolimits}
\newcommand {\mstrut}{\vphantom{\bigl(}}

\newtheorem{remark}{Замечание}
\newtheorem{predl}{Предложение}

\begin{document}

\begin{center}

\Large{{\bf БИФУРКАЦИОННЫЕ ДИАГРАММЫ НА ИЗОЭНЕРГЕТИЧЕСКИХ УРОВНЯХ ГИРОСТАТА КОВАЛЕВСКОЙ--ЯХЬЯ\footnote{Работа выполнена при финансовой поддержке гранта РФФИ № 10-01-00043.}}}

\vspace{5mm}

\normalsize

{\bf М.П.~Харламов, И.И.~Харламова, Е.Г.~Шведов}

\vspace{4mm}

\small

Волгоградская академия государственной службы

Россия, 400131, Волгоград, ул. Гагарина, 8

E-mail: mharlamov@vags.ru
\end{center}

\begin{flushright}
{\it Получено 01 июля 2010 г.}
\end{flushright}

\vspace{3mm}

\footnotesize{Предложен новый взгляд на классификацию бифуркационных диаграмм и условия существования критических движений интегрируемой задачи о движении тяжелого гиростата при условиях типа Ковалевской (случай интегрируемости Х.М.\,Яхья). Построено разделяющее множество на плоскости <<энергия-гиростатический момент>>, классифицирующее диаграммы на изоэнергетических уровнях.  Выписаны конструктивно проверяемые условия существования критических движений в терминах параметров на поверхностях, несущих листы бифуркационных диаграмм.}


\section{Постановка задачи}\label{sec1}
Случаем Ковалевской--Яхья называют задачу о движении тяжелого гиростата, главные моменты инерции которого удовлетворяют отношению 2:2:1, центр масс лежит в экваториальной плоскости, а гиростатический момент направлен по оси динамической симметрии. Подходящим выбором осей и единиц измерения уравнения движения приводятся к виду
\begin{equation}\label{neq1_1}
\begin{array}{lll}
2\dot\omega _1   = \omega _2 (\omega _3- \la)  , &
2\dot\omega _2 =  - \omega _1 (\omega _3-\la)  - \alpha _3 , &
\dot\omega _3   = \alpha _2, \\
\dot\alpha _1   = \alpha _2 \omega _3  - \alpha _3 \omega _2 , &
\dot\alpha _2   = \alpha _3 \omega _1  - \alpha _1 \omega _3 , &
\dot\alpha_3   = \alpha_1 \omega_2  - \alpha_2 \omega_1,
\end{array}
\end{equation}
где $\la \gs 0$. Фазовое пространство $\mP^5=\bbR^3_{\omega}{\times}S^2_{\alpha}$ определено в $\bbR^6$ геометрическим интегралом $|{\bs \alpha}^2|=1$. Система \eqref{neq1_1} была проинтегрирована П.В.\,Хар\-ла\-мо\-вым на трехмерных подмногообразиях, состоящих из периодических решений и их бифуркаций \cite{PVLect, PVMtt71}. Х.М.\,Яхья указал, в дополнение к классическим интегралам энергии и площадей, новый интеграл типа Ковалевской, получив полную инволютивную систему \cite{Yeh1}
\begin{equation}\label{neq1_2}
\begin{array}{l}
H = \omega _1^2  + \omega _2^2 + \ds{\frac{1}{2}}\omega _3^2 -
\alpha _1, \qquad
L = 2(\omega _1 \alpha _1  + \omega _2 \alpha _2 ) + (\omega _3+\la) \alpha _3, \\
K=(\omega_1^2-\omega^2_2+\alpha_1)^2+(2\omega_1\omega_2+\alpha_2)^2 + 2\la[(\omega_3-\la)(\omega_1^2+\omega^2_2)+2\omega_1 \alpha_3].
\end{array}
\end{equation}

Исследование множества критических точек отображения, порожденного функциями \eqref{neq1_2}, начато в \cite{Gash1,Ryab2,Gash3} и завершено в работах \cite{Gash4,Ryab1}. Как оказалось, это множество исчерпывается решениями П.В.\,Харламова. В \cite{Gash4,Ryab1} получены и уравнения бифуркационных поверхностей, т.е. связных поверхностей $\wsi_j$ в $\bbR^3(h,\ell,k)$, объединение которых содержит в себе бифуркационную диаграмму $\Sigma$ интегралов \eqref{neq1_2} как собственное подмножество\footnote{Здесь и далее постоянные первых интегралов обозначаются строчными буквами, соответствующими обозначениям самих интегралов как функций на фазовом пространстве.}. Пересечение $\Sigma_j=\Sigma \cap \wsi_j$ будем называть {\it допустимой областью} на бифуркационной поверхности $\wsi_j$.

В работах \cite{Ryab5, Ryab6, Ryab7} исследовалась эволюция сечений $\sil$ множества $\Sigma$ плоскостями $\ell=\cons$, которые с точки зрения гамильтоновой механики служат бифуркационными диаграммами приведенных систем с двумя степенями свободы, параметризованных постоянной площадей. В работе \cite{Gash2} указана топология регулярных интегральных многообразий для точек $(h,\ell,k)$ из связных компонент $\bbR^3 \backslash \Sigma$. Все перечисленные объекты и свойства зависят от одного свободного параметра -- величины гиростатического момента $\la$. В связи с
этим в работе \cite{RyabRCD} исследована зависимость диаграммы $\sila$ от {\it двух} параметров $(\ell,\la)$ и на плоскости этих параметров построено множество, при пересечении точек которого меняется тип $\sila$. Кроме того, в \cite{RyabRCD} для нулевой постоянной площадей построено множество в плоскости $(\la, h)$, классифицирующее типы графов Фоменко на трехмерных изоэнергетических уровнях, вычислены эти графы и соответствующие неоснащенные молекулы. Описание всех графов и молекул, включая случай $\ell \ne 0$, приведено в работе \cite{Gash5}. Совокупность результатов по топологическим инвариантам случая Ковалевской--Яхья подробно изложена в \cite[гл. 9]{GashDis}.
Наиболее полное исследование бифуркационных диаграмм содержится в \cite{RyabDis}, где, в частности, в терминах некоторых вспомогательных параметров решена представляющаяся аналитически наиболее сложной задача определения допустимых областей на бифуркационных поверхностях (см. \cite[\S 5.3]{RyabDis}).

Цель настоящей работы -- основываясь на перечисленных результатах, построить атлас бифуркационных диаграмм двух интегралов $G=L^2,K$ на четырехмерных изоэнергетических уровнях
$E_h(\la)=\{ H = h\} \subset \mP^5$.
Вводя вместо функции $L$ ее квадрат, мы искусственно добавляем в каждую плоскую диаграмму замыкающий ее отрезок прямой
$g=0$.
Такая постановка связана с тем, что только интеграл $G$ имеет аналог в общем случае А.Г.\,Реймана--М.А.\,Семенова-Тян-Шанского \cite{ReySem}. В частности, полученные ниже результаты являются необходимым дополнением к классификации бифуркационных диаграмм на изоэнергетических уровнях (пятимерных, ввиду отсутствия симметрии) волчка и гиростата типа Ковалевской в двойном силовом поле \cite{Kh34,KhSh34,KhND}. Таким образом, будет решена задача классификации по параметрам $h,\la$ бифуркационных диаграмм $\Sigma_h(\la)$ ограничения отображения $G{\times}K$ на подмногообразие $E_h(\la)$.

Понимая под {\it атласом} объектов полное описание классифицирующего (разделяющего) множества в пространстве параметров и возможность указания для каждой неразделяющей точки этого пространства структурно устойчивого типа самого объекта, потребуем еще наличия диалоговой компьютерной системы, которая позволяет осуществить {\it визуализацию и детализацию} разделяющего множества и объекта при интерактивном изменении параметров. Пример такой системы реализован для диаграмм волчка в двойном поле по параметрам $(h,\gamma)$, где $\gamma$ -- отношение напряженностей силовых полей \cite{Kh361}. Здесь возникает следующая проблема. Пусть $\wsi_j$ -- одна из поверхностей, несущих диаграмму $\Sigma(\beta)$, где $\beta$ -- физический параметр задачи ($\beta=\la$ для гиростата Ковалевской--Яхья и $\beta=\gamma$ для волчка в двойном поле). Предположим, что уравнения для $\wsi_j$ записаны в параметрическом виде, необходимом для построения изоэнергетических сечений, т.е. $g,k$ выражены в зависимости от $h$ и некоторой второй координаты $s$. Для построения диаграмм с помощью компьютера необходимо иметь алгоритм, позволяющий по любому~$h$ вычислить промежутки {\it фактического} изменения параметра $s$ в допустимой области. Для волчка в двойном поле такой алгоритм реализован в \cite{Kh361} путем указания допустимых областей на бифуркационных поверхностях неравенствами, в которых $h$ выступает в роли параметра. Для гиростата Ковалевской--Яхья эта проблема до сих пор не решена даже для диаграмм $\sila$, что с вычислительной точки зрения равносильно нашей постановке, поскольку в параметрических уравнениях $h$ и $g=\ell^2$ оказываются связанными линейно. В настоящей работе получена вся необходимая информация для компьютерной визуализации диаграмм гиростата Ковалевской--Яхья.

\section{Критическое множество и бифуркационные поверхности}\label{sec2}
Опираясь на результаты П.Е.\,Рябова и И.Н.\,Гашененко, представим множество критических точек отображения
$J=H{\times}G{\times}K: \mP^5 \to \bbI$ в виде трех критических подсистем -- трехмерных инвариантных подмногообразий $\mo, \mm, \mn$ в фазовом пространстве. Их образы под действием $J$ лежат на трех бифуркационных поверхностях, уравнения которых запишем в форме, вытекающей из представления Лакса \cite{ReySem} при анализе особенностей возникающей алгебраической кривой. Вводя параметр $s$ как удвоенный квадрат спектрального параметра на кривой, получим (подробности см. в работе \cite{KhND} для более общего случая):
\begin{equation}
\label{neq2_1}
\wsc :  \left\{ \begin{array}{l} \displaystyle{g=(h-\frac{\lambda^2}{2}-s)s^2,}\\[3mm]
\displaystyle{k =
1+(h-\frac{\lambda^2}{2})^2-4(h-\frac{\lambda^2}{2})s+3s^2,}
\end{array} \right.
\end{equation}
\begin{equation}
\label{neq2_2}
\wsa, \wsb : \left\{ \begin{array}{l} \displaystyle{g=\frac{1}{2}(h+\frac{\lambda^2}{2})-\lambda^2s^2-\frac{1}{4s},}\\[3mm]
\displaystyle{k = -
2\lambda^2(h-\frac{\lambda^2}{2}-2s)-\lambda^4+ \frac{1}
{4s^2}.}
\end{array} \right.
\end{equation}
Здесь $s\in \bbR$ для $\wsc$, $s <0 $ для $\wsa$ и $s > 0 $ для $\wsb$. В отличие от двойного поля, поверхность $\wsc$ имеет одну компоненту.

Движения на $\mo$ соответствуют решению, построенному в \cite{PVLect} для произвольного осесимметричного тензора инерции. Предполагая в этом решении выполненными такие же условия на физические параметры, как в системе \eqref{neq1_1}, и выбирая на $\wsc$ точку, заданную уравнениями \eqref{neq2_1} с параметрами $h,s$, будем иметь
\begin{equation}\label{neq2_3}
\begin{array}{lll}
  \ds{\omega_1=p_0,} &  \ds{\omega_2=0,} & \ds{\omega_3 = r,}\\
  \ds{\alpha_1=p_0^2+\frac{1}{2} r^2-h,} &
  \ds{\alpha_2= R, }&
  \ds{\alpha_3 = -p_0 (r-\la),}\\
\end{array}
\end{equation}
где
\begin{equation}\notag
\begin{array}{l}
   \ds{p_0^2 = h-\frac{\la^2}{2}-s,} \\
   \ds{R^2=-\frac{1}{4}r^4-(2p_0^2-h)r^2+2 \la p_0^2 r+1-(p_0^2-h)^2-p_0^2\la^2.}
\end{array}
\end{equation}

Движения на $\mm$ и $\mn$ описываются решением П.В.\,Хар\-ла\-мова, найденным в работе \cite{PVMtt71}.  При заданных $h,s$ и $g$, удовлетворяющих \eqref{neq2_2}, введем следующие обозначения
\begin{equation}\notag
\begin{array}{c}
   \vk ^2 = g+\la^2 s^2, \quad \ro^2=1-\ds{\frac{2 \vk^2}{s}},  \quad
   \Fun^2=\ds{\frac{1}{2}}\left[\bigl(X+\ds{\frac{\la}{\vk}}\bigr)^2+\bigl(\ro Y +\ds{\frac{\sqrt{g}}{s \vk}}\bigr)^2-1\right],\\
   (X,Y)=\left\{ \begin{array}{ll}
   (\cos \sigma, \sin \sigma), & \ro^2 \gs 0 \\
   (\cosh \sigma, \ri \sinh \sigma), & \ro^2 < 0
   \end{array}\right.
\end{array}
\end{equation}
Здесь $\ri$ -- мнимая единица, $\sigma$ -- вспомогательная переменная. Многообразия $\mm,\mn$ описываются уравнениями
\begin{equation}\label{neq2_4}
\begin{array}{lll}
  \ds{\omega_1=-\frac{\sqrt{g}} {s}- \vk \ro Y,} &
  \ds{\omega_2=-\ro \sqrt{s}\, \Fun,} &
  \ds{\omega_3 = \la+2 \vk X,}\\[2mm]
  \ds{\alpha_1=\frac{\la s X+\sqrt{g}\ro Y}{\vk} -2\vk^2Y^2 ,} &
  \ds{\alpha_2=-2 \vk Y \sqrt{s} \, \Fun, } &
  \ds{\alpha_3 = \frac{\sqrt{g}X-\la s \ro Y }{\vk},}
\end{array}
\end{equation}
а динамика задана уравнением $\dot \sigma^2  = \sgn(\ro^2)\, s \, \Fun^2$. При этом $s<0$ для $\mm$ и $s>0$ для $\mn$.

\section{Бифуркационные диаграммы критических подсистем и существование движений}\label{sec3}
Особым случаям {\it внутри} критических подсистем отвечают особенности поверхностей \eqref{neq2_1}, \eqref{neq2_2}. Геометрия этих поверхностей достаточно хорошо изучена в работах \cite{Ryab2,Gash4,Ryab1,RyabDis}. Нам понадобятся уточнения, связанные с условиями существования критических движений. Имеется три кривые, по которым $\wsc$ пересекается с объединением $\wsa \cup \wsb$ трансверсально. В части, принадлежащей $\Sigma$, обозначим их через $\delta_1,\delta_2,\delta_3$ (необходимые формулы будут даны ниже). Кроме этого, поверхность $\wsc$ касается $\wsb$ по кривой, пересечение которой с $\Sigma$ обозначим через $\gan$.

\begin{remark}[1]
Указанные кривые и возникающие на них узловые точки, будучи одними и теми же объектами в $\bbI$, получают различное представление в координатах на поверхностях $\wsi_j$. Несмотря на это, их образы в $(s,h)$-плоскостях будем для наглядности обозначать одинаково.
\end{remark}

По определению $g\gs 0$, поэтому в границы допустимых областей всегда входят линии пересечения $\wsi_j$ с плоскостью $g=0$. Обозначим соответствующие кривые через $\gamma_j$ ($j=1,2,3$).

Введем следующую систему обозначений. Представляя кривую ${\gamma_j\subset\wsi_j}$ в соответствующей $(s,h)$-плоскости, рассмотрим ее уравнение $h=h(s)$. Если в заданных пределах изменения $h$ решение относительно $s$ единственно, то обозначим его через $\xi_j(h,\la)$. Если же этих решений два, то обозначим их через $\xi^-_j(h,\la) < \xi^+_j(h,\la)$. Каждая из кривых $\delta_i$ может отображаться в различных $(s,h)$-плоскостях. Проведем сечение образа $\delta_i$ в $(s,h)$-плоскости, отвечающей поверхности $\wsi_j$, на заданном уровне $h$. Если на рассматриваемой ветви кривой точка пересечения единственна, то ее \mbox{$s$-}координату обозначим через $\eta_{ij}(h,\la)$. Если же таких точек две, то их \mbox{$s$-}координаты обозначим через $\eta_{ij}^-(h,\la)< \eta_{ij}^+(h,\la)$.

Рассмотрим систему $\mo$ с точки зрения условий существования вещественных движений.
\begin{predl}[1]
При заданных $s,h$ вещественные решения \eqref{neq2_3} существуют тогда и только тогда, когда $p_0^2\gs 0$ и $R^2(r)\gs 0$ для некоторого $r\in \bbR$.
\end{predl}

Бифуркациям решений \eqref{neq2_3} отвечают точки $(s,h)$, в которых либо $p_0=0$, либо многочлен $R^2(r)$ имеет кратный корень. Выбирая, следуя работе \cite{Gash4}, этот корень в качестве параметра, на дискриминантном множестве будем иметь
\begin{equation}\label{neq3_1}
    \begin{array}{c}
      h=\varphi_{\pm}(r)=\ds{\frac{1}{2}\bigl[ r (\la-r) \pm \frac{2r-\la}{r-\la}D\bigr]},\quad
      s=\psi_{\pm}(r)=\ds{\frac{1}{2}\bigl[  \la (r-\la) \pm D\bigr]},\\[2mm]
      D=\sqrt{\mstrut r^2(r-\la)^2+4} \gs 0.
    \end{array}
\end{equation}
Из предложения~1 получим следующее утверждение.
\begin{predl}[2]
Бифуркационная $(s,h)$-диаграмма критической системы $\mo$ состоит из следующих множеств:
\begin{equation}\notag
    \begin{array}{ll}
      \gaa: \quad h=s+\ds{\frac{\la^2}{2}} , & s \gs -1 - \ds{\frac{\la^2}{2}};\\[2mm]
      \delta_1: \quad h=\varphi_-(r), & s=\psi_-(r), \quad r \in [0,\la);\\[2mm]
      \delta_2: \quad h=\varphi_+(r), & s=\psi_+(r), \quad r \in (-\infty,0];\\[2mm]
      \delta_3: \quad h=\varphi_+(r), & s=\psi_+(r), \quad r \in (\la,+\infty).
    \end{array}
\end{equation}
При этом внешними границами области существования движений служат кривые $\gaa, \delta_1$ и $\delta_3$.
\end{predl}

Диаграмма системы $\mo$ показана на рис. 1,\,{\it a}. Здесь и далее на аналогичных рисунках звездочкой отмечены связные компоненты дополнения диаграммы, не входящие в допустимую область.

\begin{figure}[ht]
\centering
\includegraphics[width=100mm, keepaspectratio]{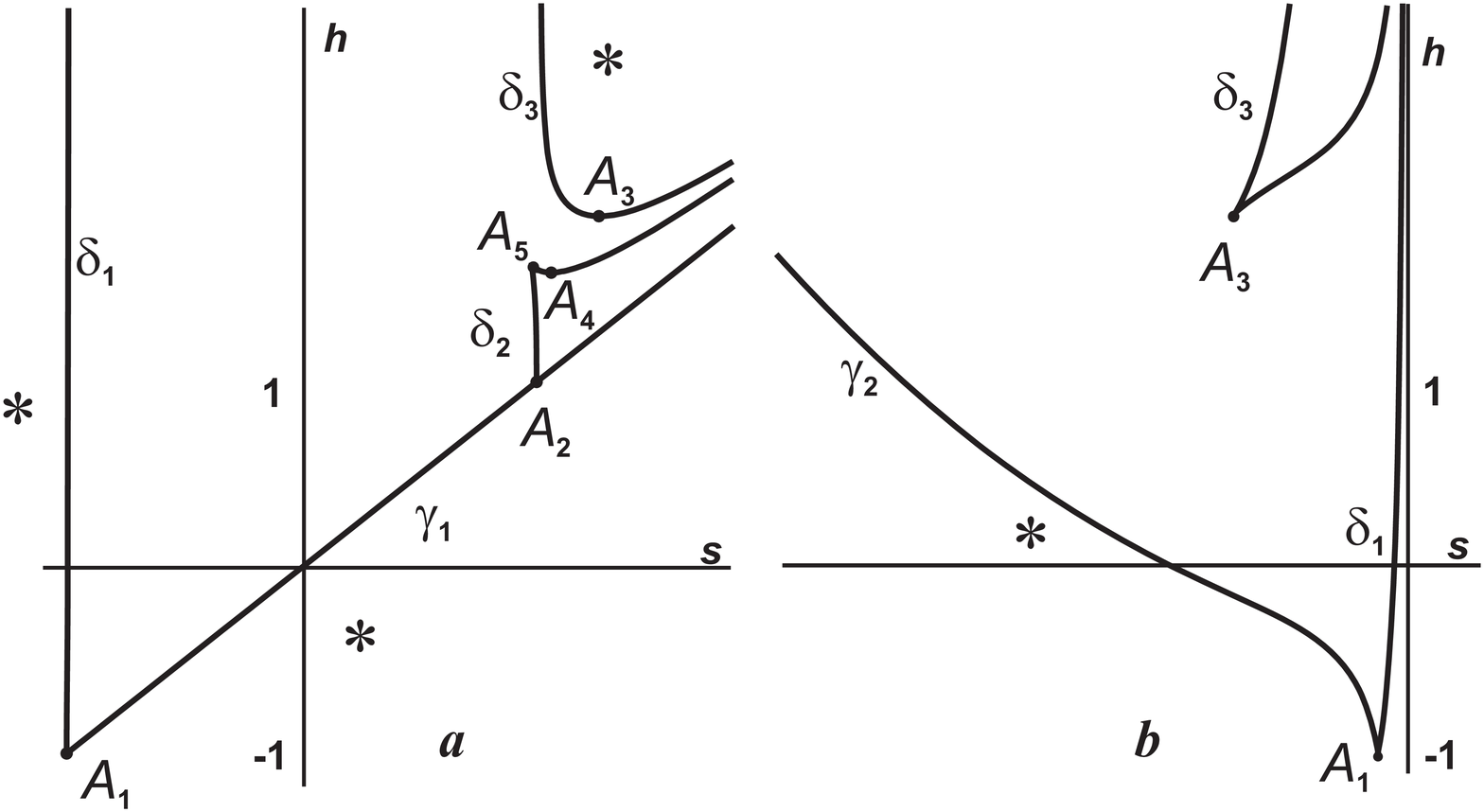}
\caption{Допустимые области: $a)$ для $\mo $,  $b)$ для $\mm $.}\label{fig1}
\end{figure}

Точки $A_1, A_2$ отвечают глобальным критическим значениям энергии
\begin{equation}\label{neq3_2}
    A_{1,2}: \quad h=\mp 1, \quad \la \gs 0.
\end{equation}

На кривой $\delta_1$ зависимость $h(r)$ монотонна, поэтому для любого ${h\gs =-1}$ существует единственное решение
\begin{equation}\label{neq3_3}
    r_- = r_-(h,\la) \in [0,\la)
\end{equation}
уравнения $\varphi_-(r)=h$. Очевидно, по принятым обозначениям, $\eta_{11}=\psi_-(r_-)$.

Перестройки $h$-сечений допустимой области происходят при прохождении величиной $h$, в дополнение к значениям \eqref{neq3_2}, экстремумов на кривых $\delta_2,\delta_3$, т.е. точек, обозначенных через $A_3 - A_5$. Уравнение $\varphi'_+(r)=0$ распадается на два
\begin{eqnarray}
& &(2r-\la)(r-\la)-D=0, \label{neq3_4}\\
& &r(2r-\la)(r-\la)+\la D=0. \label{neq3_5}
\end{eqnarray}
Вычисляя $\psi'_+(r)$, убеждаемся, что условие \eqref{neq3_4} задает экстремумы на гладких участках -- точки $A_3,A_4$, а условие \eqref{neq3_5} определяет точку возврата $A_5$.
Из \eqref{neq3_4} получаем уравнение
\begin{equation}\label{neq3_6}
    (r-\la)^3(3r-\la)-4=0,
\end{equation}
которое имеет ровно два корня, причем верхний всегда больше $\la$, поэтому это единственный экстремум $h$ на $\delta_3$ (точка $A_3$), а нижний оказывается неположительным лишь при $\la \ls \sqrt{2}$. В последнем случае имеем экстремум $h$ на $\delta_2$ (точка $A_4$). Явной функции $h(\la)$ в этих точках построить не удается. Поступим следующим образом. Положим
в уравнении \eqref{neq3_6}
\begin{equation}\label{neq3_7}
    x=\la - r.
\end{equation}
Из \eqref{neq3_6} выразим $\la(x)$, а из \eqref{neq3_7} найдем $r(x)$ и подставим в выражение для $h$ на кривых $\delta_2,\delta_3$. Получим
\begin{equation}\notag
    \la = \ds{\frac{3x^4-4}{2x^3}}, \qquad r = \ds{\frac{x^4-4}{2x^3}}, \qquad h = \ds{\frac{3}{8} x^2 +\frac{2}{x^6}},
\end{equation}
где $x\in [-\sqrt[4]{4/3}\, ,0)$ для $A_3$ и $x\in [\sqrt[4]{4/3} \, , \sqrt{2}\,]$ для $A_4$. Поскольку $\la \gs 0$, можно записать единую параметризацию зависимости $h$ и $\la$, положив $u=x^2$:
\begin{equation}\label{neq3_9}
    A_{3,4}: \quad h_{3,4}=\ds{\frac{3}{8}u+\frac{2}{u^3}}, \quad \la^2_{3,4}=\ds{\frac{(3u^2-4)^2}{4u^3}}, \quad u \in (0,2].
\end{equation}
При $h>h_3$ уравнение $\varphi_+(r)=h$ имеет ровно два вещественных корня в области $r\in (\la,+\infty)$. Их подстановка в функцию $\psi_+$ дает зависимости
$\eta^-_{31} (h,\la)$ и $\eta^+_{31}(h,\la)$ на монотонных участках $\delta_3$. Очевидно, $\eta^-_{31} < \eta^+_{31} < \xi_1=h-{\la^2}/{2}$.

Уравнение \eqref{neq3_5} вместе с уравнением кривой $\delta_2$ дает в точке возврата
\begin{equation}\label{neq3_10}
    A_{5}: \quad h_{5}=\ds{\frac{1}{4}\left[(4+\la^{4/3})^{3/2}-\la^{2/3}(6+\la^{4/3}) \right] }, \quad \la \gs 0.
\end{equation}

\begin{predl}[3]
Допустимая область на поверхности $\wsc$ описывается следующим образом:
\begin{equation}\notag
    \begin{array}{lll}
      h \in (-\infty,-1) & \Rightarrow & s \in \varnothing;\\
      h \in [-1,h_3] & \Rightarrow & s \in [\eta_{11}, h-\ds{\frac{\la^2}{2}}] ;\\
      h \in (h_3,+\infty) & \Rightarrow & s \in [\eta_{11}, \eta^-_{31}] \cup [\eta^+_{31}, h-\ds{\frac{\la^2}{2}}].
    \end{array}
\end{equation}
\end{predl}

Подчеркнем, что $\eta_{11}, \eta^-_{31}, \eta^+_{31}$ -- однозначно определенные и эффективно вычисляемые функции от $h,\la$.

Для систем $\mm,\mn$ условие $g\gs 0$ равносильно
\begin{equation}\label{neq3_11}
    h \gs h_{\rm min} = \ds{\frac{1-\la^2 s+4 \la^2 s^3}{2s}},
\end{equation}
а из уравнений \eqref{neq2_4} следует, что $\sgn (\ro^2) =\sgn (Y^2)$ и $\sgn (\ro^2) =\sgn (s \Fun^2)$.
Пусть $Y_*=Y$, если $\ro$ вещественно, и $Y_*=\ri Y$, если $\ro$ чисто мнимое. Тогда в плоскости $(X,Y_*)$ кривая $\Gamma_0$, заданная тождеством $X^2+Y^2=1$, представляет собой окружность или гиперболу, в то время как кривая $\Gamma_1$, заданная уравнением $\Fun^2(X,Y) = 0$, при всех $\ro^2 \ne 0$ есть эллипс.
Отметим также следующие из определений равенства
\begin{equation} \label{neq3_12}
\begin{array}{c}
    \ro^2 = \ds{\frac{\htan-h}{s}}, \quad  \htan = \ds{\frac{1-\la^2 s+2 s^2}{2s}},\quad
    h_{\rm min} - \htan = - s( 1-2\la^2s).
\end{array}
\end{equation}
Здесь $h=\htan(s)$ -- зависимость на кривой $\gan$ касания поверхностей $\wsc$ и $\wsb$.

Из \eqref{neq3_11}, \eqref{neq3_12} получаем следующее утверждение.
\begin{predl}[4]
Для существования вещественных решений \eqref{neq2_4} при заданных $s,h$ необходимо и достаточно выполнение условия \eqref{neq3_11} и следующих условий:

$1)$ при $s<0$ окружность $\Gamma_0$ и эллипс $\Gamma_1$ имеют общую точку$;$

$2)$ при $s>0, \; \ro^2 \gs 0$ окружность $\Gamma_0$ не лежит целиком строго внутри области, ограниченной эллипсом $\Gamma_1$$;$

$3)$ при $s>0, \; \ro^2 < 0$ гипербола $\Gamma_0$ и эллипс $\Gamma_1$ имеют общую точку.
\end{predl}

Отсюда следует, что бифуркации решений по $s,h$ происходят в случаях касания кривых второго порядка $\Gamma_0$ и $\Gamma_1$. Обозначая в точке касания $\omega_3=r$ (здесь это константа), получим выражения
\begin{equation}\label{neq3_13}
    \begin{array}{c}
      h=\varphi_{\pm}(r),\quad
      s=\theta_{\pm}(r)=\ds{\frac{r-\la}{4\la}\bigl[r(r-\la) \mp D\bigr]},
    \end{array}
\end{equation}
где $D, \varphi_{\pm}$ определены в \eqref{neq3_1}. Это совпадение не случайно, поскольку найденные значения отвечают точкам трансверсального пересечения поверхности $\wsc$ с $\wsa, \wsb$. Выражения для $s$ различны, так как они определяют одну из компонент нормали к соответствующей поверхности в общей точке.

Анализируя условия предложения~4 в областях, на которые полуплоскость $s<0$ делится кривыми \eqref{neq3_13}, с учетом условия \eqref{neq3_11} приходим к следующему результату.

\begin{predl}[5]
Бифуркационная $(s,h)$-диаграмма критической системы $\mm$ состоит из следующих множеств:
\begin{equation}\notag
    \begin{array}{ll}
      \gab: \quad h=h_{\rm min}(s) , & s \ls -\ds{\frac{1}{2}};\\[2mm]
      \delta_1: \quad h=\varphi_-(r), & s=\theta_-(r), \quad r \in [0,\la);\\[2mm]
      \delta_3: \quad h=\varphi_+(r), & s=\theta_+(r), \quad r \in (\la,+\infty).
    \end{array}
\end{equation}
Внешними границами области существования движений служат $\gab$ и $\delta_1$.
\end{predl}

Поскольку зависимость \eqref{neq3_3} на $\delta_1$ уже известна, получаем
$\eta_{12}=\theta_-(r_-)$.
На кривой $\gab$ имеем
\begin{equation}\notag
    h_{\rm min}'(s) = \ds{\frac{1}{2s^2}}(8\la^2 s^3-1).
\end{equation}
Поэтому при отрицательных $s$ уравнение $h=h_{\rm min}(s)$ имеет единственное решение
$s = \xi_2(h,\la)$. Тогда из предложения~5 получим следующее утверждение.
\begin{predl}[6]
Допустимая область на поверхности $\wsa$ описывается следующим образом:
\begin{equation}\notag
    \begin{array}{lll}
      h \in (-\infty,-1) & \Rightarrow & s \in \varnothing;\\
      h \in [-1,+\infty) & \Rightarrow & s \in [\xi_2, \eta_{12}].
    \end{array}
\end{equation}
\end{predl}
И здесь функции $\eta_{12}(h,\la), \xi_2(h,\la)$ эффективно вычисляются.

Диаграмма системы $\mm$ показана на рис. 1,\,{\it b}. Новых разделяющих зависимостей между $h$ и $\la$ не появляется, поскольку узловые точки $A_1$ и $A_3$ уже учтены в системе $\mo$.
\begin{remark}[2]
Отметим двойственность, типичную для случая, когда ребро возврата одной бифуркационной поверхности попадает на другую. Возникающий след в параметрах одной поверхности дает экстремум на гладкой кривой, а в параметрах другой порождает точку возврата. Здесь это видно для точки $A_3$, но ниже проявится и для точек $A_4,A_5$.
\end{remark}

Для системы на $\mn$ к уже известным уравнениям бифуркаций решений~\eqref{neq2_4} добавляется разделяющий случай $\ro^2=0$. Из последнего равенства~\eqref{neq3_12} определяются значения $s$, при которых эта кривая лежит в области \eqref{neq3_11}.
\begin{predl}[7]
Бифуркационная $(s,h)$-диаграмма критической системы $\mn$ состоит из следующих множеств:
\begin{equation}\notag
    \begin{array}{lll}
      \gan: & h=\htan(s), & 0< s \ls \ds{\frac{1}{2\la^2}};\\[2mm]
      \delta_2: & h=\varphi_+(r), & s=\theta_+(r), \quad r \in (-\infty,0]; \\[2mm]
      \gac: & h=h_{\rm min}(s) , & s \in I(\la),
    \end{array}
\end{equation}
где
\begin{equation}\notag
    I(\la)= \left\{
    \begin{array}{ll}
      (0,+\infty), & \la^2 \ls 8/(3\sqrt{3}) \\
      (0, s_*] \cup [s^*, +\infty), & 8/(3\sqrt{3}) \ls \la^2 \ls 2\\
      (0, s_*] \cup [1/2, +\infty), & \la^2 \gs 2
    \end{array}\right.,
\end{equation}
$s_*(\la)<s^*(\la)$ -- абсциссы точек касания кривых $\delta_2$ и $\gac$, существующих при значениях ${\la^2 \gs 8/(3\sqrt{3})}$.

Внешними границами допустимой области служат:

$1)$ кривая $\gac$ в указанных пределах$;$

$2)$ кривая $\gan$ в пределах
\begin{equation}\notag
    s \in  \left\{
    \begin{array}{ll}
      (0, 1/(2 \la^{2/3})] , & \la^2 \ls 1/(2\sqrt{2}) \\
      (0, \sqrt{1+\la^4}-\la^2], & \la^2 \gs 1/(2\sqrt{2})
    \end{array}
    \right. ;
\end{equation}

$3)$ кривая $\delta_2$ в пределах
\begin{equation}\notag
   s\in
   \left\{
   \begin{array}{ll}
      \, [s_*, s^*]  , & 8/(3\sqrt{3}) \ls \la^2 \ls 2 \ \\
      \, [s_*, 1/2]    , & \la^2 \gs 2
   \end{array}
   \right. .
\end{equation}
\end{predl}

Кривая $\delta_2$ заканчивается на кривой $\gac$ при $r=0$, $s=1/2$. Это -- отмеченная ранее точка $A_2$. Уравнение для точки возврата кривой $\delta_2$ совпадает с \eqref{neq3_4}, поэтому это точка $A_4$. Экстремум $h$ на $\delta_2$ -- это точка $A_5$ (замечание~2).

Кривая $\delta_2$ имеет с кривой $\gan$ точку касания $A_6$ при $s=\sqrt{1+\la^4}-\la^2$, ${r=-\la}$ для всех $\la >0$. Точка их трансверсального пересечения $A_7$, в которой ${s=1/(2\la^{2/3})}$, $r=\la - 1/\sqrt[3]{\la}$, существует при $0<\la \ls 1$. При проходе кривой $\gan$ в сторону увеличения $s$ точка $A_6$ встречается первой при $8\la^4<1$, затем $A_6$ и $A_7$ меняются местами. Верхняя граница $s$ на $\gan$ определяется из последнего уравнения \eqref{neq3_12} точкой $A_8$ пересечения с кривой $\gac$. При условии $\la^2 < 1/\sqrt{2}$ на кривой $\gan$ реализуется минимум $h$, равный $\sqrt{2}-\la^2/2$. Соответствующая точка $A_9$ лежит левее $A_7$ при $\la^2 < 1/(2\sqrt{2})$, а при увеличении $\la$ она находится правее $A_6$ и на допустимую область уже не влияет. Отметим еще минимум $h$ на кривой $\gac$ (точка $A_{10}$), равный $(3\la^{2/3}-\la^2)/2$.

На рис.~2 показаны первые три варианта бифуркационной $(s,h)$-диаграм\-мы при возрастании $\la$: 1) ${0<\la^2<1/(2\sqrt{2})}$;
2) ${1/(2\sqrt{2})<\la^2<1}$; 3) ${\la^2> 1}$. Проверка условий предложения 4 показывает, что движения невозможны в области, ограниченной двумя бесконечными отрезками кривых $\delta_2$ и $\gan$ до точки пересечения $A_7$ в первом случае и до точки касания $A_6$ при остальных $\la$. Как отмечалось, это свойство впервые доказано (в иных терминах и параметрах) в работе \cite{RyabDis}. Обсуждаемая недопустимая область на рис.~2 указана звездочкой со стрелками.

\begin{figure}[ht]
\centering
\includegraphics[width=130mm, keepaspectratio]{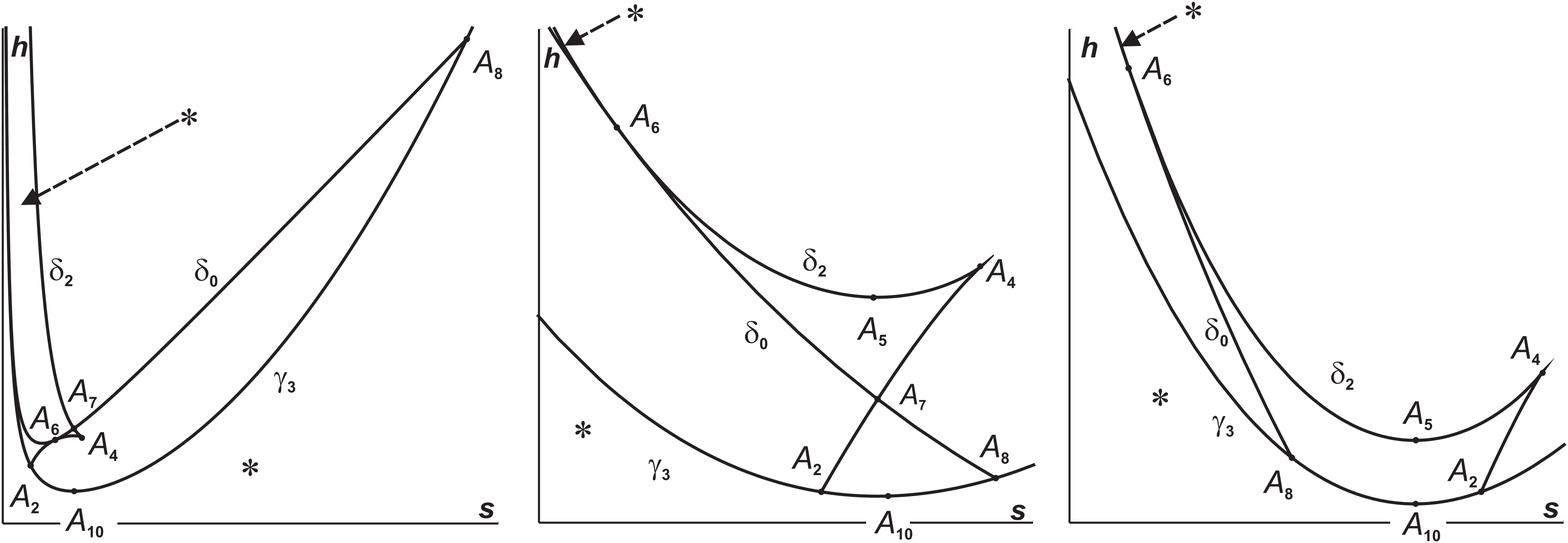}
\caption{Допустимая область для $\mn$.}\label{fig2}
\end{figure}

\begin{remark}[3]
Строго говоря, пересечение кривой $\gan$ не всегда сопровождается топологическими бифуркациями решений, но заведомо меняется их аналитическое представление \eqref{neq2_4}. Как следует из результатов \cite{KhND}, эта кривая также отвечает за вырождение индуцированной симплектической структуры на соответствующем четырехмерном критическом подмногообразии фазового пространства $TSO(3)$ нередуцированной системы.
\end{remark}

\begin{figure}[ht]
\centering
\includegraphics[width=100mm, keepaspectratio]{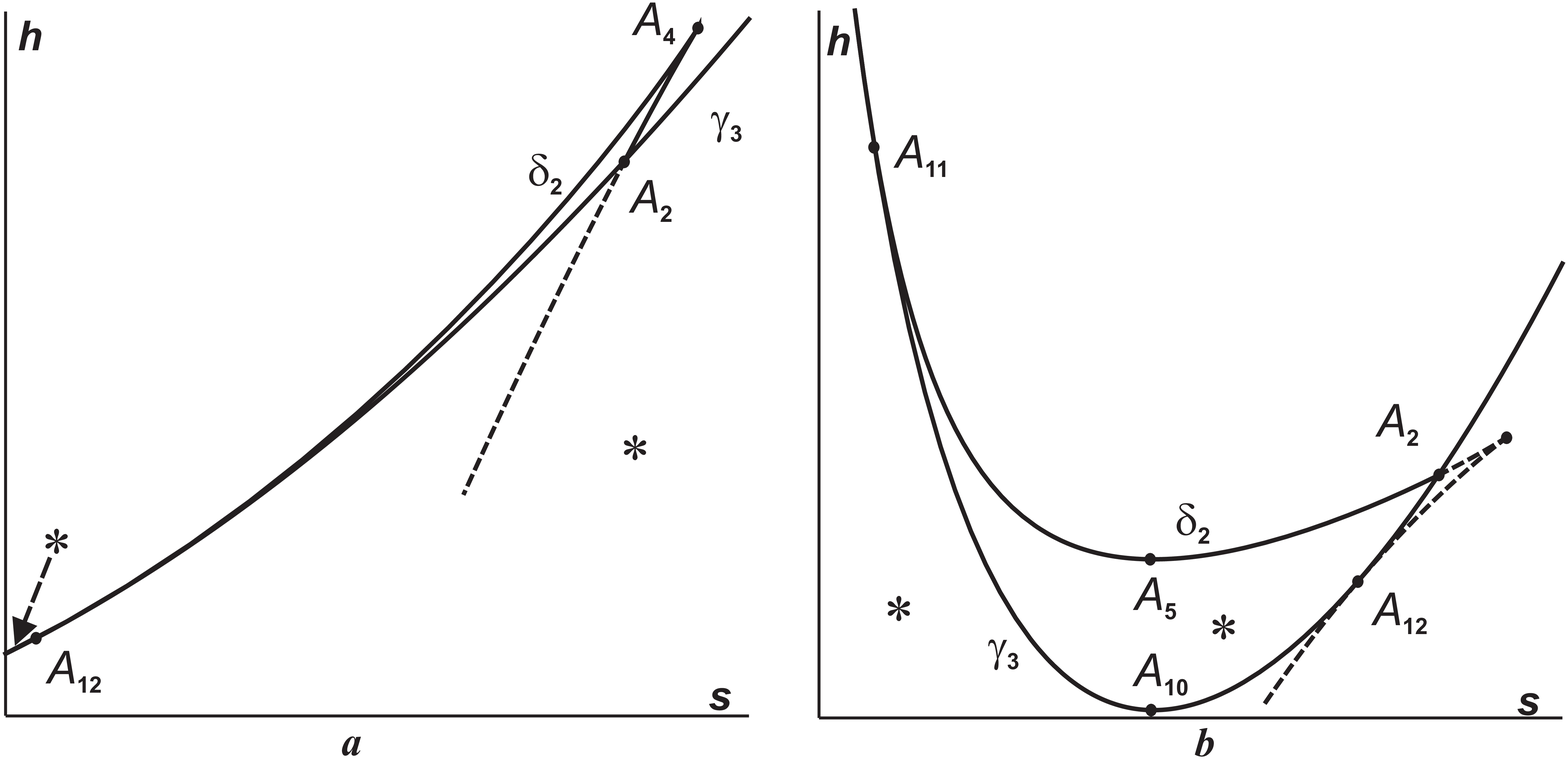}
\caption{Эволюция особых точек в окрестности $\la^2=2$.}\label{fig3}
\end{figure}

Приведенными тремя случаями видоизменения $(s,h)$-диаграммы на $\mn$ не исчерпываются. Точка возврата $A_4$ кривой $\delta_2$
существует лишь при ${\la^2 < 2}$. При $\la^2 = 2$ она сливается с $A_2$ и после этого исчезает. При $r<0$ имеем
$$
\ds{\frac{\varphi_+(r)-h_{\rm min}(\theta_+(r))}{\theta_+(r)}=-\frac{\la r}{(r-\la)^2}\left[\la(r-\la)+D\right]^2} \gs 0,
$$
поэтому другие возможные общие точки $\gac$ и $\delta_2$ оказываются точками касания и подчинены уравнению
\begin{equation}\label{neq3_15}
(\la-r)^3(\la+r)- 4 =0.
\end{equation}
Оно имеет вещественные корни лишь при $\la^2 \gs 8/(3\sqrt{3})$. В случае строгого неравенства таких корней два, нижний всегда отрицателен, а верхний отрицателен лишь при $\la^2<2$. Поэтому если ${8/(3\sqrt{3})< \la^2 < 2}$, то между точками $A_6,A_5$ кривая $\delta_2$ имеет еще две точки касания $A_{11}, A_{12}$ с кривой $\gac$.
Соответствующие значения $s$ в предложении 7 обозначены через $s_*,s^*$. При $\la^2=2$ точка $A_{12}$ сливается с $A_5$ и $A_2$ и при $\la^2>2$ уходит из допустимой области вместе с точкой возврата (см. рис.~3).
При $g=0$ между точками $A_8$ и $A_2$ имеем $\ro^2=1-2\la^2 s <0$, поэтому совместность уравнений \eqref{neq2_4} равносильна существованию решений системы неравенств
$$
X^2 \gs 1, \qquad Q(X)=2\la^2 s^3 (X^2-1)+2s X+1 \ls 0.
$$
Между точками $A_{11}$ и $A_{12}$ дискриминант $Q(X)$ отрицателен, а при $\la^2 > 2$ между точками $A_{11}$ и $A_2$ оба корня $Q(X)$ по модулю меньше 1.
Поэтому на этих интервалах кривой $\gac$ вещественных решений не существует. Следовательно, при ${\la^2 > 8/(3\sqrt{3})}$ нижней границей допустимой области является соответствующий участок кривой $\delta_2$ (от $A_{11}$ до $A_{12}$ при $\la^2 \ls 2$ и от $A_{11}$ до $A_2$ при $\la^2>2$).
Неявную зависимость $h$ и $\la$ для $A_{11,12}$ можно параметризовать, используя для уравнения \eqref{neq3_15} замену \eqref{neq3_7}.

В разделяющее множество на плоскости $(h,\la)$ для атласа диаграмм $\Sigma_h(\la)$, в дополнение к найденным ранее кривым  \eqref{neq3_2}, \eqref{neq3_9}, \eqref{neq3_10}, необходимо включить кривые, отвечающие зависимостям между $h$ и $\la$ в новых особых точках:
\begin{equation}\label{neq3_16}
    \begin{array}{ll}
      A_6: & h_6 = \ds{\frac{3}{2}} \sqrt{1+\la^4} -\la^2, \qquad  \la \gs 0;\\
      A_7: & h_7 = \ds{-\frac{\la^2}{2}} + \la^{2/3}+ \ds{\frac{1}{2\la^{2/3}}}, \qquad  0 < \la \ls 1;\\
      A_8: & h_8 = \ds{\frac{1+\la^4}{2\la^2}}, \qquad  \la > 0;\\
      A_9 : & h_9=\sqrt{2}-\la^2/2, \qquad  0 \ls \la \ls \sqrt{1/(2\sqrt{2})};\\
      A_{10}: & h_{10} = \ds{\frac{1}{2}(3\la^{2/3}-\la^2)}, \qquad  0 \ls \la \ls \sqrt{8/(3\sqrt{3})};\\[1.5mm]
      A_{11,12}: & h_{11,12} = \ds{\frac{1}{8} x^2 +\frac{2}{x^2}-\frac{2}{x^6} }, \quad \la = \ds{
   \frac{1}{2} x +\frac{2}{x^3}}, \qquad  x \in [\sqrt{2},+\infty ).
    \end{array}
\end{equation}

Для явного указания интервалов изменения $s$ в зависимости от $\la,h$ обозначим через $\zeta^- < \zeta^+$ решения уравнения $h=\htan(s)$ при $h \gs h_9$
$$
\zeta^{\pm}(h,\la)= \ds{\frac{1}{2}}\left[h+\ds{\frac{\la^2}{2}} \pm \sqrt{\left(h+\ds{\frac{\la^2}{2}}
\right)^2-2} \, \right]
$$
и напомним, что положительные решения уравнения $h=h_{\rm min}(s)$ при ${h \gs h_{10}}$ обозначаются $\xi^-_3 < \xi^+_3$. Здесь и далее $h_j$ определены в соответствии с \eqref{neq3_16}.

При $\la^2 \ls 1/(2\sqrt{2})$ на участке кривой $\delta_2$ при $h \gs h_7$, а при остальных $\la$ -- на участке $h\gs h_6$ имеем однозначную зависимость $s=\eta_{23}(h,\la)$.
При $\la^2 \gs 8/(3\sqrt{3})$ на монотонно убывающем бесконечном участке кривой $\delta_2$ до пересечения с точкой $A_{11}$ имеем решение $s=\eta^-_{23}(h,\la)$, а от точки $A_{12}$ до точки $A_2$ -- решение $s=\eta^+_{23}(h,\la)$.
Получаем следующее утверждение.
\begin{predl}[8]
Допустимая область на поверхности $\wsb$ такова: при $0< \la^2 \ls 1/(2\sqrt{2})$
\begin{equation}\notag
    \begin{array}{lll}
      h \in (-\infty,h_{10}) & \Rightarrow & s \in \varnothing;\\
      h \in [h_{10},h_9] & \Rightarrow & s \in [\xi^-_3, \xi^+_3] ;\\
      h \in (h_9,h_7) & \Rightarrow & s \in [\xi^-_3, \zeta^-] \cup [\zeta^+, \xi^+_3]; \\
      h \in (h_7,+\infty) & \Rightarrow & s \in [\xi^-_3, \zeta^-] \cup [\eta_{23}, \xi^+_3],
    \end{array}
\end{equation}
при $ 1/(2\sqrt{2}) \ls \la^2 \ls 8/(3\sqrt{3})$
\begin{equation}\notag
    \begin{array}{lll}
      h \in (-\infty,h_{10}) & \Rightarrow & s \in \varnothing;\\
      h \in [h_{10},h_6] & \Rightarrow & s \in [\xi^-_3, \xi^+_3] ;\\
      h \in (h_6,+\infty) & \Rightarrow & s \in [\xi^-_3, \zeta^-] \cup [\eta_{23}, \xi^+_3];
    \end{array}
\end{equation}
при $ 8/(3\sqrt{3}) \ls \la^2 \ls 2 $
\begin{equation}\notag
    \begin{array}{lll}
      h \in (-\infty,h_5) & \Rightarrow & s \in \varnothing;\\
      h \in [h_5,h_{12}] & \Rightarrow & s \in [\eta^-_{23}, \eta^+_{23}] ;\\
      h \in [h_{12},h_{11}] & \Rightarrow & s \in [\eta^-_{23}, \xi^+_3] ;\\
      h \in [h_{11},h_6] & \Rightarrow & s \in [\xi^-_3, \xi^+_3] ;\\
      h \in (h_6,+\infty) & \Rightarrow & s \in [\xi^-_3, \zeta^-] \cup [\eta_{23}, \xi^+_3];
    \end{array}
\end{equation}
при $ \la^2 \gs 2 $
\begin{equation}\notag
    \begin{array}{lll}
      h \in (-\infty,h_5) & \Rightarrow & s \in \varnothing;\\
      h \in [h_5,h_2] & \Rightarrow & s \in [\eta^-_{23}, \eta^+_{23}] ;\\
      h \in [h_2,h_{11}] & \Rightarrow & s \in [\eta^-_{23}, \xi^+_3] ;\\
      h \in [h_{11},h_6] & \Rightarrow & s \in [\xi^-_3, \xi^+_3] ;\\
      h \in (h_6,+\infty) & \Rightarrow & s \in [\xi^-_3, \zeta^-] \cup [\eta_{23}, \xi^+_3].
    \end{array}
\end{equation}
\end{predl}

Для пороговых значений $\la$, естественно, можно пользоваться любым из двух подходящих представлений.

Таким образом, представлена вся информация по классификации бифуркационных диаграмм на изоэнергетических уровнях гиростата Ковалевской--Яхья. Разделяющее множество $\mathcal{C}$ на плоскости $(h,\la)$ задано уравнениями \eqref{neq3_2}, \eqref{neq3_9}, \eqref{neq3_10}, \eqref{neq3_16}. Отметим, что пересечение $\mathcal{C}$ с осью $\la=0$ дает значения
$
h=-1, \, 0,\, 1,\, \sqrt{2},\, \frac{3}{2},\, \sqrt{3},\, 2,
$
классифицирующие изоэнергетические диаграммы классического волчка Ковалевской (см., например, предельный случай в~\cite{KhSh34}).

Программы символьных вычислений позволяют вывести на экран компьютера множество $\mathcal{C}$ и, выбирая интерактивно точку в какой-либо из областей, получить при необходимости детализацию окрестности этой точки. Позиционируя окончательное значение пары параметров $(h,\la)$ в более мелком масштабе, согласно предложениям~3, 6, 8 строим диаграмму $\Sigma_h(\la)$ по формулам \eqref{neq2_2}, \eqref{neq2_1} с уже вполне определенными промежутками изменения параметра~$s$. В результате получаем компьютерную систему, реализующую построение искомого электронного атласа.

\end{document}